\documentclass[12pt]{iopart}

\usepackage{iopams}  
\usepackage{graphicx}
\begin{document}

\title[Unveiling the Higgs mechanism]{Unveiling the Higgs mechanism to students}

\author{Giovanni Organtini}

\address{"Sapienza", Universit\`a di Roma \& INFN-Sez. di Roma, Roma, P.le A. Moro 2 I-00185}
\ead{giovanni.organtini@roma1.infn.it}
\begin{abstract}
In this paper we give the outline of a lecture given to undergraduate students aiming at understanding why physicists are so much interested in the Higgs boson. The lecture has been conceived for students not yet familiar with advanced physics and is suitable for several disciplines, other than physics. The Higgs mechanism is introduced by semi-classical arguments mimicking the basic field theory concepts, assuming the validity of a symmetry principle in the expression of the energy of particles in a classical field. The lecture is divided in two parts: the first, suitable even to high--school students, shows how the mass of a particle results as a dynamical effect due to the interaction between a massless particle and a field (as in the Higgs mechanism). The audience of the second part, much more technical, consists mainly of teachers and university students of disciplines other than physics.
\end{abstract}

\submitto{\EJP}
\maketitle

\section{Introduction}
On july, 4$^{th}$, 2012, both the ATLAS and CMS experiments at LHC announced the discovery of a new
particle, believed to be the Higgs boson.

The emphasis given by all the media to the discovery, makes more people asking for a deeper understanding of the nature and the properties of the Higgs boson. However, explaining how the introduction of a new particle can give mass to other particles is extremely difficult without quantum field theory.

A very common way to explain how this is possible, makes use of a famous CERN cartoon~\cite{higgs}, based on suggestions from Prof. David Miller, in which the Higgs mechanism is described in terms of scientists at a conference (representing the Higgs field) that meet a famous colleague (representing a massless particle). As a result they struggle for speaking with him, making it moving through the room with some {\em inertia}, leading to the appearance of its mass as a dynamical effect.

In this lecture we provide a more formal explanation of this fact, introducing an {\em ad hoc} symmetry, mimicking what happens in quantum field theory (QFT)~\cite{qft}. As in QFT weak interactions break the so-called gauge symmetry, leading to inconsistencies in the theory, in our approach we observe that the classical expression of the energy of a particle in a field seems to obey a particular symmetry. Such a symmetry is broken by Special Relativity, as in QFT the gauge symmetry is broken by weak interactions. Introducing a new field, we can keep the original symmetry and, at the same time, maintain the validity of Special Relativity, analogously with what happen in QFT.

The symmetry required by us comes from esthetic arguments, indeed, and of course is not strongly motivated. Though arbitrary, this symmetry principle mimics very well what happens in quantum field theory and allows an explanation, at classical level, of a purely quantistic effect.
It must be clear, then, that what follows must be regarded just as an artifact used to show how the breaking of a symmetry can lead to some dynamical appearance of the inertial mass. For a consistent theory, see~\cite{Bernstein}.

The lecture outlined below has been tuned for high--school students. As a consequence, we adopted a simplified formalism, avoiding integrals and derivatives, and choosing a particular system to analyze: a charged particle inside a parallel plate capacitor. It can be shown, however, that the considerations made for such a system can be easily extended to any system of particles, inside a region with an arbitrary, conservative, field. 

\section{The energy of a particle in a field}
Consider a particle with mass $m$ and charge $q$ in an electric field $E$, whose potential is $V$. The potential energy of the particle can be written as

\begin{equation}\label{eq:classical_energy_3}
U=qV\,.
\end{equation}
Electric fields themselves carry energy and we can compute the energy contained in a given volume ${\cal V}$ in vacuum as

\begin{equation}
U = \frac{\epsilon_0}{2}{E^2}{\cal V}\,,
\end{equation}
where $\epsilon_0$ is the dielectric constant of the vacuum.
Consider now a parallel plate capacitor, with a charged particle between the plates. The total energy contained in the capacitor volume ${\cal V}$ can then be computed as the sum of the energy stored in the capacitor and the one of the charged particle:

\begin{equation}\label{eq:einvolume}
U = qV + \frac{\epsilon_0}{2}E^2{\cal V}\,.
\end{equation}
Here we assumed, for simplicity, that the capacitor volume is filled with vacuum. 
In order to make the expression of the energy independent on the particular geometry, we can write the energy density, i.e. the energy per unit volume, as

\begin{equation}\label{eq:edensity}
u = \frac{U}{\cal V} = \frac{qV}{\cal V} + \frac{\epsilon_0}{2}E^2\,.
\end{equation}
We observe that the energy density in a volume ${\cal V}$ is given by the sum of two terms: one is due to the interaction of a charge $q$ with an electric field, whose potential is $V$; the other is due to the interaction of a field $E$ with itself (giving rise to the term proportional to $E^2$). It seems that the energy density expression obey a simple symmetry: all the terms are built multiplying a field by (another) field or the potential of such a field times a characteristic of a particle that acts as a source of that field (the charge $q$).

Note that this is true for all the known conservative forces: the energy contained in a given volume is always written as a sum of two kinds of terms:

\begin{enumerate} 
\item products of two fields;
\item products of the corresponding potentials times sources of the fields.
\end{enumerate}
The observed symmetry is broken as soon as we introduce special relativity, for which everyone knows that the energy of a particle of mass $m$ at rest is $mc^2$. As a consequence we need to add such a term to the expression of $u$ in all the equations written above, for each particle in the considered volume. For example, equation~(\ref{eq:edensity}) becomes

\begin{equation}
u =  \frac{U}{\cal V} = \frac{qV}{\cal V} + \frac{\epsilon_0}{2}E^2 + \frac{mc^2}{\cal V}\,.
\end{equation}
The minimal possible energy is reached when there are no particles nor fields in a volume, i.e. when $E=0$ and $m=0$. The minimum of the energy corresponds to an {\em empty} volume. We are used to call vacuum this condition. We could alternatively define the vacuum as a condition in which the energy reaches its lowest value; on the other hand the two definitions coincide.

The relativistic term breaks the symmetry described above, because the last term is not in the form of a product of a potential times a source or in the form of the product of two fields. Despite the fact that there is no need to maintain the observed symmetry, it is somewhat disturbing to have an extra term depending on $m$, also because there is no term like $q$ times some universal constant. That means that $m$ and $q$ cannot be considered both just as coupling constants. There must be something fundamentally different between $m$ and $q$, for which $m$ alone contributes to the energy of a particle, while $q$ does not.

Let's then formulate a somewhat arbitrary principle, according to which the energy contained in a given volume is always given in terms of a sum of products of at least two {\em objects}: a particle and a field or two fields. On the other hand there cannot be interaction if we have only one particle or one non auto--interacting field in a volume. If we assume such a principle, the term $mc^2$ cannot be part of the energy expression, because it is not in the above form.

\section{Introducing the Higgs field}
Let's rewrite the energy density $u$ in a volume {\cal V}, ignoring, for the time being, the relativistic term:

\begin{equation}
u = \frac{U}{\cal V} = \frac{q}{\cal V}V+\frac{\epsilon_0}{2}E^2\,. 
\end{equation}

Suppose now that there is some field $\phi$, not yet discovered, such that it couples with both matter and fields. Assuming that the energy is the sum of all possible products of two fields or potentials times sources, we must write

\begin{equation}\label{eq:modified_energy}
u = \frac{q}{\cal V}V + \frac{\epsilon_0}{2}E^2 + \frac{a}{\cal V}\Phi + gE\phi + g'\phi^2
\end{equation}
$\Phi$ being the potential of the field $\phi$.
Equation~(\ref{eq:modified_energy}) was obtained writing down all the possible particle--field (the first and the third terms) and field--field pairs, each multiplied by a coupling constant. In fact, the first two terms are, respectively, of the forms

\begin{enumerate}
\item\label{f1} [constant][$1/{\cal V}$][some function of a field], and
\item\label{f2} [constant][a field][a field].
\end{enumerate}
Constants are $q$ and $\epsilon_0/2$, respectively. They are called {\em coupling constants} and give the intensity of the coupling between a particle and a field, or two fields. $1/{\cal V}$ in some way represents matter: it can be thought as the {\em density} of the particle in the volume. To these terms we added

\begin{itemize}
\item a term of the form (\ref{f1}), where the constant is $a$, the particle is represented as $1/{\cal V}$, as above, and the function of the new field $\phi$ is $\Phi$;
\item a term of the form (\ref{f2}), where the constant is $g$ and the fields are $E$ and $\phi$;
\item a term of the form (\ref{f2}), where the constant is $g'$ and the fields are both equal to $\phi$\,.
\end{itemize}
There are no other possible combinations. To simplify notation one can also denote with ${\cal P}$ the factor $1/{\cal V}$, with $F_p$ a function of the field $F$, representing its potential, $c_i$ a coupling constant, so that, having just one particle and one field $E$, the energy is written as

\begin{equation}
u = c_1{\cal P}E_p + c_2EE\,,
\end{equation}
where, if $E$ is the electric field, $c_1=q$, $c_2=\epsilon_0/2$, $E_p=V$ and $V$ is a function of both $E$ and $x$, i.e. $V=V(E,x)$. With three {\em objects}, a particle and two fields, we have

\begin{equation}
u = c_1{\cal P}E_p + c_2EE + c_3{\cal P}\phi_p + c_4E\phi + c_5\phi\phi\,,
\end{equation}
and the symmetry is evident.

\begin{figure}
\begin{center}
\includegraphics[width=0.4\textwidth]{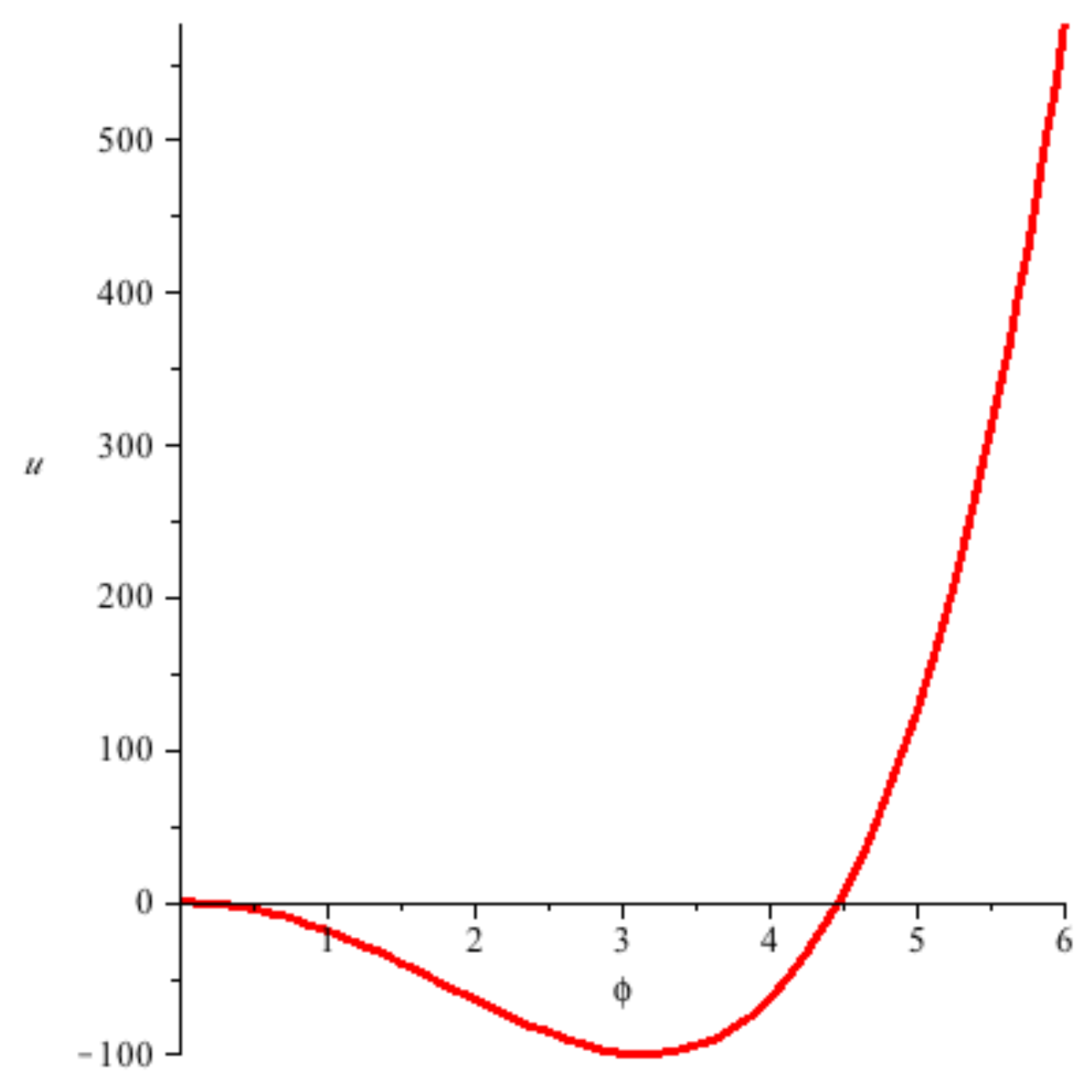}
\caption{\label{fig:higgs1d}Graph of the function $u=-g'\phi^2+\phi^4$. Here we have put $g' = 20$ and the minimum is attained at $\phi = \sqrt{10} \simeq 3.2$.}
\end{center}
\end{figure}

It is easy to show that adding a term like $\phi^4$ to the energy density, the latter reaches its minimum for a field $\phi \ne 0$. In fact, consider the function $u=-g'\phi^2+\phi^4$, shown in Figure~\ref{fig:higgs1d}. Keeping only positive solutions, $u$ reaches its minimum for $\phi=\phi_0=\sqrt{g'/2}$. The lowest possible energy is no more reached when all fields are zero, but in a condition for which the field $\phi \ne 0$.

We now change our definition of vacuum. If we define the vacuum as the condition in which the energy reaches its minimum, we see that the vacuum is something in which there is some field $\phi=\phi_0$. A completely empty space, i.e. a space for which $E=m=\phi=0$ does not necessarily coincide with vacuum, as long as its energy is larger than that of the vacuum.

If we rewrite equation~(\ref{eq:modified_energy}) making it manifest that $g'<0$ and adding an extra term $\phi^4$ we have

\begin{equation}\label{eq:modified_energy_2}
u = \frac{q}{\cal V}V + \frac{\epsilon_0}{2}E^2 + \frac{a}{\cal V}\Phi + gE\phi - |g'|\phi^2+\phi^4\,.
\end{equation}
This expression is identical to the previous one and can be interpreted in the same way, but for the presence of the term $\phi^4$ that, however, does not break the symmetry of the energy density expression: the energy density is still a sum of products of at least two objects. With this energy density, vacuum is the condition in which $E=0$ and $\phi=\phi_0$, i.e. vacuum is such that there is some field in the volume.

\section{The Higgs boson}
Let's write $\phi=\phi_0 + \eta$. Correspondingly, its potential $\Phi$ can be written as $\Phi = \Phi_0 + \Phi_1$, $\Phi_0$ being the vacuum potential and $\Phi_0 \ne 0$. Since $\Phi_0$ is the potential of the field $\phi$ in vacuum, it must be constant everywhere.

Under this hypothesis, the energy density contained in our capacitor is then

\begin{equation}\label{eq:modified_energy_2}
u = \frac{q}{\cal V}V + \frac{\epsilon_0}{2}E^2 + \frac{a}{\cal V}\left(\Phi_0 + \Phi_1\right) + gE(\phi_0 + \eta) - |g'|\phi^2+\phi^4\,.
\end{equation}
The current expression for the energy density is in fact very interesting. Let's analyze it term by term, starting from the third one, given that the first two terms are actually the classical ones.

We observed that a mass term is something of the form of a constant times ${1/{\cal V}}$.  In equation~(\ref{eq:modified_energy_2}), a mass term appears as the interaction of a (massless) particle with the vacuum potential. In fact, the mass term of the particle is exactly the one predicted by Einstein if we put

\begin{equation}
mc^2 = a\Phi_0 \Rightarrow m = \frac{a}{c^2}\Phi_0\,.
\end{equation}
Note that, being $\Phi_0$ constant, masses are constants. We say that the mass term is due to an interaction because its form is the same of $qV/{\cal V}$, representing the interaction of a particle $1/{\cal V}$ with a field $E$, whose potential is $V$, with coupling constant $q$. By analogy, $a\Phi_0/{\cal V}$ represents the interaction of the same particle with the field $\phi$, whose potential is $\Phi$, with coupling constant $a$.
An extra interaction term appears, as

\begin{equation}
\frac{a}{\cal V}\Phi_1\,.
\end{equation}
This term represents the interaction of a particle with a field $\eta$, called the Higgs field, whose potential is $\Phi_1$, with coupling constant 

\begin{equation}
a=\frac{mc^2}{\Phi_0}\,,
\end{equation}
i.e. the intensity of the interaction is proportional to the particle mass. A discharged capacitor, with no charges nor electric field inside, is not necessarily {\em empty}. It contains at least the vacuum field $\phi_0$, whose potential is $\Phi_0$.

Because of the term $gE\eta$, the field $E$ must interact with the Higgs field, too. However, if $g \ne 0$, $E$ acquire a mass due to the term $gE\phi_0$, in contrast with experiments. The term $gE\phi_0$ represents a mass for the field, analogously to the term $a\Phi_0/{\cal V}$: mass terms are those for which {\em a particle} ($1/\cal V$) or a field is multiplied by a constant. We are then forced to put $g=0$.

Expanding the $\phi$ field in the last two terms gives rise to the following terms:

\begin{equation}
-|g'|\phi_0^2-2|g'|\phi_0\eta-|g'|\eta^2+(higher\; order\; terms)
\end{equation} 
The first three terms are, respectively, a constant (irrelevant), a mass term for the Higgs field ($-2|g'|\phi_0\eta$) and the auto--interaction of the Higgs field ($-|g'|\eta^2$). Higher order terms are just combinations of these terms, representing more complex interactions that can be neglected. For example, the term $4\phi_0^3\eta$ represents a process in which the vacuum $\phi_0$ interacts with itself (giving rise to a term proportional to $\phi_0^2$) together with a process in which the Higgs field interacts with the vacuum, making a factor $\phi_0\eta$ to appear.

Apparently, then, with the hypothesis that vacuum is not empty and requiring that the energy density in a region of the Universe is written as a sum of products of at least two {\em objects}, we can explain the mass terms in the energy, at the price of introducing a new field (the Higgs field), interacting, in principle, with both matter and fields and auto--interacting. Moreover we have to pay the price of accepting a mass term for the electric field $E$, unless we believe that $g=0$.

The mechanism outlined above is enough for most of the students. It is simple enough to be understood and explains quite naturally why we need the Higgs boson and how it works. The Higgs mechanism consists in introducing a new field such that the mass terms in the energy expression appears dynamically, analogously to any other term in such an expression. 

We repeat here that there is no reason, in classical physics, to require that the energy expression must show the symmetry imposed by us. We can justify the requirement only by {\em aesthetic} arguments that are rather weak, indeed.
On the other hand, with this principle we are able to mimic what happens in quantum field theory, where a similar symmetry principle is needed to preserve gauge invariance (that, on the contrary, has been experimentally proven). 

The lecture ends here. That is enough to understand the Higgs mechanism. However, for those most motivated we also provide a simple enough explanation of the spontaneous symmetry breaking mechanism, giving rise to mass terms for the weak field, maintaining the electric field massless.

\section{Weak interactions}
All the fields studied in an undergraduate course are massless: the electric field $E$, the magnetic field $B$ and the gravitational field $G$. Since there is no mass term associated to those fields, then we are encouraged to put $g=0$ in our symmetric expression of the energy density. That appears unnatural because there is no fundamental reason for that.

It turns out that, in fact, at least a massive field exists: the one responsible for weak interactions. A massive field is a field that cannot propagate at infinite distances like $E$, $B$ or $G$. In fact, suppose we have a particle of mass $M$ at rest, producing a given field. Because of conservation of mass, it can only produce a massless field. In fact, if it produces a massive field with mass $\mu$, then it must loose some mass and become a particle of mass $M-\mu$. This is impossible, unless it happens in such a way that we can never measure it. Remember: Physics has to do only with what can be measured. If something cannot be measured in any way, whether it exists or not is irrelevant for physics.

It turns out that in quantum mechanics, the {\em uncertainty principle} holds, according to which we can never measure together energy and time with infinite precision. If $\Delta E$ and $\Delta t$ are, respectively, the uncertainty in the measurement of energy and time, the uncertainty principle states that

\begin{equation}
\Delta E \Delta t \ge \frac{h}{2\pi}
\end{equation}
where $h$ is the Planck's constant. Note that the uncertainty principle has nothing to do with the fact that every measurement is affected by an error; it is not related to the technology of the instruments. Even if every measurement is always affected by some measurement, we can always imagine that, at least in principle, we can build energy and time measurement devices such that their uncertainties are such that their product is lower than $h/2\pi$, even if not null. The uncertainty principle holds independently on the technology used. 

Thanks to the uncertainty principle, we can violate energy conservation for time intervals $\Delta t$ lower than

\begin{equation}
\Delta t \le \frac{h}{2\pi\Delta E}\,.
\end{equation}
In fact these violations are unobservable. A particle of mass $M$, then, can produce a massive field with mass $\mu$, provided that it lasts for at most

\begin{equation}
\Delta t \le \frac{h}{2\pi\mu c^2}\,,
\end{equation}
being the conservation of energy violated by $Mc^2-(Mc^2-\mu c^2)=\mu c^2$. A massive field cannot travel at the speed of light, but for sure it cannot exceed it. Then, suppose that it propagates at the speed of light: it cannot go farther than

\begin{equation}
L = c\Delta t \le \frac{h}{2\pi\mu c}\,,
\end{equation}
giving rise to limited range interactions, such as weak interactions. Weak interactions, responsible for nuclear reactions, must then be represented by massive fields. 

\section{Spontaneous symmetry breaking}
Taking into account weak interactions, we are lead to add extra terms in the expression of the energy density in a given volume. In our classical expression we may have to add something like

\begin{equation}
u = u_0 + \frac{w}{\cal V}{\cal Z} + \frac{\zeta_0}{2}Z^2
\end{equation}
where $u_0$ is the right side of equation~(\ref{eq:classical_energy_3}), $w{\cal Z}$ is the product of a {\em weak charge} $w$ times the potential $\cal Z$ of the field $Z$, representing some interaction of $w$ with $Z$. $\zeta_0Z^2/2$ is analogous to $\epsilon_0E^2/2$ and represents some auto--interaction of the weak field ($\zeta_0$ is some constant to be determined by experiments).

The two terms representing the auto--interaction of fields, $\epsilon_0E^2/2+\zeta_0Z^2/2$, can be thought as arising from the product of just one field with two components $\sqrt{\epsilon_0}E$ and $\sqrt{\zeta_0}Z$. Think the two components as the components of a bi--dimensional vector $\mathbf{V}$ on a horizontal plane. The scalar product of the vector $\mathbf{V}$ times itself, gives 

\begin{equation}
\frac{\mathbf{V}\cdot \mathbf{V}}{2} = \frac{\epsilon_0}{2}E^2 + \frac{\zeta_0}{2}Z^2\,.
\end{equation}
Accordingly, we can define a bi--dimensional vector $\mathbf{v}=\left(q, w\right)$ representing coupling constants. The full expression for the energy density, without mass terms, is then simplified as

\begin{equation}
u = \frac{\mathbf{v}}{\cal V}\cdot\mathbf{P} + \frac{\mathbf{V}^2}{2}\,,
\end{equation}
$\mathbf{P}$ being a vector formed with the two potentials $\mathbf{P} = \left(V, {\cal Z}\right)$. The energy density formulation respects our symmetry: it is formed as a sum of products of at least two {\em objects} of the type specified above.
Now, let's introduce the Higgs field as in the previous section. Since the energy density is now given in terms of products of bi--dimensional vectors, the vacuum field must be a vector too: $\mbox{\boldmath{$\phi$}}=(\phi, \phi')$ and the full energy density reads

\begin{equation}\label{eq:final_equation}
u = \frac{\mathbf{v}}{\cal V}\cdot\mathbf{P} + \frac{\mathbf{V}^2}{2}\ + \frac{\mathbf{a}}{\cal V}\cdot \mbox{\boldmath{$\Phi$}} +  g\mathbf{V}\cdot\mathbf{\mbox{\boldmath{$\phi$}}} -|g'|\mbox{\boldmath{$\phi$}}^2+\mbox{\boldmath{$\phi$}}^4\,,
\end{equation}
$\mathbf{a}$ being a bi--dimensional vector whose coordinates are the coupling constants between particles and the fields $\phi$ and $\phi'$.
The mass terms arise if we write

\begin{equation}
\mbox{\boldmath{$\phi$}} = (\phi_0 + \eta, \phi'_0 + \eta')
\end{equation}
so that the fourth term gives 

\begin{equation}
gE{\sqrt{\epsilon_0}}\left(\phi_0 + \eta\right) + g{Z}{\sqrt{\zeta_0}}\left(\phi_0' + \eta'\right)\,.
\end{equation}
Note that the energy density for an {\em empty} region (i.e. a region without particles, nor electric or weak field) is now a function of two variables $\phi$ and $\phi'$ and it represents a surface in three dimensions, obtained rotating the shape of $u=-g'\phi^2+\phi^4$ around the $u$--axis (see Fig.~\ref{fig:higgs1d}). 
As usual, mass terms are those proportional to the {\em size of the vacuum} that, in this case, is $\phi_0$ along the $\phi$--axis and $\phi_0'$ along the $\phi'$--axis. We then have two mass terms:

\begin{equation}
m_Ec^2 = g{E}{\sqrt{\epsilon_0}}\phi_0
\end{equation}
and

\begin{equation}
\mu c^2 = g{Z}{\sqrt{\zeta_0}}\phi_0'\,.
\end{equation}
Contrary to our experience, both fields acquire a mass. However, any rotation of the vector $\mbox{\boldmath{$\phi$}}$ around the $u$--axis, leads to the same energy and we can then arbitrarily choose one of the infinite vacuum states represented by any vector lying on the horizontal plane whose length is $\sqrt{|g'|/2}$. In particular we can choose a state for which the coordinates of the vacuum vector are $\left(0, \phi^*\right)$. In other words, we rotate the reference frame until the vacuum vector lies along one of the axis.

With this choice, equivalent to any other choice, the mass terms become, respectively, $m_E=0$ and $\mu c^2 = gZ\phi^*\sqrt{\zeta_0}$. We then have a massless field and a massive one, with mass $\mu$. Impressive!

The rest is straightforward. The third term in equation~(\ref{eq:final_equation}) gives a mass term for the particle as well as an interaction term with a Higgs boson.  Let's write it explicitly:

\begin{equation}
a\mathbf{V}\cdot \mbox{\boldmath{$\phi$}}=a\frac{q\sqrt{\epsilon_0}}{\cal V}\left(\phi_0 + \eta\right) + a\frac{w\sqrt{\zeta_0}}{\cal V}\left(\phi_0' + \eta'\right)\,.
\end{equation}
Since we choose a particular vacuum state, we have $\phi_0 = 0$, $\phi_0' = \phi^*$, then

\begin{equation}\label{eq:mass_as_interaction}
a\mathbf{V}\cdot \mbox{\boldmath{$\phi$}}=a\frac{q\sqrt{\epsilon_0}}{\cal V}\eta + a\frac{w\sqrt{\zeta_0}}{\cal V}\left(\phi^* + \eta'\right)\,,
\end{equation}
and the term $aw\sqrt{\zeta_0}\phi^*=mc^2$ represents the mass term for the matter particle. An interaction appears between the matter particle and the two fields $\eta$ and $\eta'$ as the two additional terms in equation~(\ref{eq:mass_as_interaction}).

Now remember that the vacuum consists in a condition in which the vector $\mbox{\boldmath{$\phi$}}$ has coordinates $\mbox{\boldmath{$\phi$}}_{vac} = (0, \phi^*)$ and the fields $\eta$ and $\eta'$ arise when the vector $\mbox{\boldmath{$\phi$}}$ is slightly different from $\mbox{\boldmath{$\phi$}}_{vac}$, that lies on one of the horizontal axis in a three dimensional reference frame. Imagine the arrow of this vector that moves around this position: the first coordinate oscillates around zero, while the second coordinate fluctuates between $\phi^*-\eta'$ to $\phi^*+\eta'$. On average, then, even if $\eta\ne 0$, the energy arising from the interaction between matter and the first Higgs field $\eta$ is null: its contribution to the energy is sometimes positive, sometimes negative. Only the energy due to the interaction between the particle and the second Higgs field $\eta'$ is always positive and, on average, different from zero. We can then neglect the contribution of the term proportional to $\eta$ and the interaction term between the Higgs field and particles becomes

\begin{equation}
a\frac{w\sqrt{\zeta_0}}{\cal V}\eta'\,.
\end{equation}
At the same time, since $\eta \simeq 0$,  the interaction between $E$ and $Z$ with the Higgs field becomes

\begin{equation}
g{Z}{\sqrt{\zeta_0}}\eta'\,.
\end{equation}
The Higgs field does not couple with the electric field, but do that with the weak one. It is exactly the same result obtained in quantum field theory, where the Higgs boson couples only with massive fields\footnote{In fact a coupling with massless fields, such as photons, is possible via virtual quantum loops.}.

\section{Summary}
In summary, assuming that the energy density in a given volume must be given by a sum of products between a sum of terms related to at least two components of the Universe (particles or fields), and introducing some auto--interacting field defining the vacuum state as the state with the minimum possible energy, we conclude that

\begin{itemize}
\item mass terms for particles arise naturally as the energy of the interaction of the particles with the vacuum;
\item mass terms appear also for fields, but observing that there are infinite vacuum states we can exploit the symmetry of the vacuum in such a way that it is always possible to redefine it as the one that makes one or more of the possible fields massless;
\item we obtain a massless field and a massive field, as experimentally observed;
\item there must be some extra massive field, interacting with particles and other massive fields, including itself. Such a field is the Higgs field. 
\end{itemize}
The validity of the assumptions made above allows us to explain why particles have masses that contribute to the energy density and why there are massive fields with finite range. We don't believe that this is a coincidence. We then believe in the existence of the Higgs field, that can be detected at particle accelerators such as LHC. This Higgs field shows up as a massive particle, like the weak field, that, because of that, has a finite lifetime. The higher the mass, the shorter the lifetime. After a very short time $\Delta t$ such that $\Delta t m_H c^2 \le h/2\pi$, $m_H$ being the Higgs mass, it must disappear. It then couple with a particle--antiparticle pair, created from the vacuum, in which is said to decay. It is this pair that is detected in the experiments.

\section{Remarks}
For the teachers it must be said that the way in which we have outlined the spontaneous symmetry breaking mechanism is not, actually, very rigorous. Indeed, in quantum field theory, one of the degrees of freedom of the Higgs doublet can always be set to zero because of gauge invariance. On the other hand, the fact that we can arbitrarily choose the vacuum state such that one of its components vanish works exactly the same in quantum field theory. 

Despite the fact that we do worked out the mechanism in a classical world using somewhat arbitrary arguments to justify our choices, the result is very similar to that obtained in the Standard Model of electroweak interactions and helps in understanding the fundamental difference between a real coupling constant like the electric charge $q$ from something different like the mass $m$.

We deliberately choose to write the energy density of a massive particle in an electric field, rather than in a gravitational field, because the latter is not yet completely understood in terms of quantum field theory. Gravity is difficult to manage in quantum physics because the concepts of position, velocity and acceleration are not precisely defined in this theory. Moreover, the fact that $m$ is at the same time a dynamical property due to the interaction of particles with the Higgs field, and the coupling constant to the gravitational field makes it peculiar with respect to other interactions. It must be clear that the Higgs mechanism explains the existence of the inertial mass. Why the inertial mass is equivalent to the gravitational mass is a yet unresolved problem.

The magnetic field $B$ gives rise to the same results of the electric field $E$, but there are complications due to the fact that $B$ has a vector potential and is not worth to include it in high--school.

Perhaps, the weaker point in the above discussion is the assumption that the term $g\phi_0E$ is analogous to $a\Phi_0/{\cal V}$ to justify the existence of a mass term for a field. It can be proven rather rigorously that this is the case writing $\Phi$ as $\Phi=\Phi_0+g(\phi,x)$, where $g(\phi,x)$ is a suitable function of the field $\phi$ and position $x$. On the other hand $\Phi$ is a potential, i.e. a state function of the field that can depend, then, only on the field itself and the position. The minimum energy is obtained for $\Phi=\Phi_0$ when $\phi=\phi_0$, it follows that $g(\phi_0, x)=0$. Let's expand $g(\phi,x)$ in series, around $\phi_0$:

\begin{equation}
g(\phi,x) \simeq g(\phi_0, x) + (\phi - \phi_0)\left.{\frac{\partial g}{\partial \phi}}\right|_{\phi=\phi_0}=(\phi - \phi_0)\left.{\frac{\partial g}{\partial \phi}}\right|_{\phi=\phi_0}\,.
\end{equation}
Writing $\phi=\phi_0 + \eta$ and substituting in the mass term:

\begin{equation}
\frac{a}{\cal V}\Phi = \frac{a}{\cal V}\left[\Phi_0 + \eta\left.{\frac{\partial g}{\partial \phi}}\right|_{\phi=\phi_0}\right]\,.
\end{equation}
For dimensional reasons $[\Phi_0]=[\phi]$, then $\Phi_0 = \phi_0 + \Delta\phi$, with $\Delta\phi$ constant. $\Phi_0$ has the same dimensions of $\phi_0$ and is constant. Since we can always add arbitrary constants to energy we are free to put $\Delta\phi=0$, so that $\Phi = \phi_0 + \alpha\eta$, with $\alpha$ dimensionless. Due to the fact that $\Phi$ is a scalar by definition, $\phi$ is a scalar field, too. 

\section{Conclusion}
We show an original derivation of the Higgs mechanism in classical physics, making it understandable by high--school students, not used to manage quantum fields, operators and group theory. We believe that the outlined theory, though somewhat arbitrary in few points, is very effective in popularization of such an important result in quantum field theory.

\ack
I'm grateful to Prof. M. Testa for reading the manuscript and for useful discussions.

\end{document}